\newcommand{\ab}{\mathbf{a}}

\newcommand{\eb}{\mathbf{e}}

\newcommand{\nb}{\mathbf{n}}

\newcommand{\rb}{\mathbf{r}}

\newcommand{\Nb}{\mathbf{N}}

\newcommand{\Tb}{\mathbf{T}}

\documentclass[]{article}
\usepackage{amssymb,amsfonts,amsthm,bm,graphicx,amsmath,float,color}
\restylefloat{figure}
\usepackage[ansinew]{inputenc}

\begin{document}
\title{Protein-induced membrane curvature changes membrane tension}

\author{Padmini RANGAMANI \footnote{\textbf {Corresponding author}: padmini.rangamani@berkeley.edu} \ and Kranthi K. MANDADAPU\footnote{\textbf {Corresponding author}: kranthi@berkeley.edu} \ and George OSTER\footnote{\textbf {Corresponding author}: goster@berkeley.edu}
\\[0.05in]
\sl University of California, Berkeley, CA 94720, USA
\rm \normalsize}
\maketitle


\tableofcontents

\begin{abstract}
\noindent \normalsize{Adsorption of proteins onto membranes can alter the local membrane curvature. This phenomenon has been observed in biological processes such as endocytosis, tubulation and vesiculation. However, it is not clear how the local surface properties of the membrane, such as membrane tension, change in response to  protein adsorption. In this paper, we show that the classical elastic model of lipid membranes cannot account for simultaneous changes in shape and membrane tension due to protein adsorption in a local region, and a viscous-elastic formulation is necessary to fully describe the system. Therefore, we develop a viscous-elastic model for inhomogeneous membranes of the Helfrich type. Using the new viscous-elastic model, we find that the lipids flow to accommodate changes in membrane curvature during protein adsorption. We show that, at the end of protein adsorption process, the system sustains a residual local tension to balance the difference between the actual mean curvature and the imposed spontaneous curvatures. This change in membrane tension can have a functional impact in many biological phenomena where proteins interact with membranes.} 

\end{abstract}


\section{Introduction}
Adsorption of proteins onto a membrane can induce curvature of the bilayer \cite{Kukulski2012}. There are many examples where this process is necessary for cellular function. For example, the binding of coat proteins that initiate endocytosis \cite{Kishimoto2011,Buser2013}, scaffolding of the membrane by protein complexes such as eisosomes \cite{Walther2006,Karotki2011}, and curvature sensing and modulation by the BAR domain family of proteins \cite{Stachowiak2012,Kishimoto2011,Kukulski2012,Arkhipov2008}. In many of these instances, the proteins bind cooperatively to form large complexes such that diffusion of the protein complexes is negligible \cite{Walther2006,Arkhipov2008,Valentine2011},  and so these complexes are essentially stationary. The intrinstic structure of these proteins induce the lipid bilayer to curve so as to  accommodate the shape of the protein \cite{Olivera2011,Kishimoto2011,Arkhipov2008,Buser2013}. Changes in the membrane shape in response to protein binding is documented in many ultrastructure characterization studies \cite{Frost2008,Buser2013}. However, it is not clear how adsorption of proteins on the surface alters the properties of the lipid bilayer, nor how  the lipids flow to accommodate the shape change of the membrane. In this study, we seek to answer these questions using continuum mathematical models and numerical simulations of the lipid bilayer under the influence of curvature-inducing proteins. 

Protein adsorption can alter the lateral pressure profile on the bilayer and induce tubulation \cite{Stachowiak2012,Lipowsky2012}. In \cite{Stachowiak2012,Zhao2013}, the authors use GFP-tagged proteins that adsorbed onto the bilayer. These proteins raise the lateral surface pressure and induce tubules to grow from lipid vesicles. In \cite{Lipowsky2012}, the Helfrich potential is used to show that adsorbed proteins give rise to spontaneous surface tension. However, the coupling between the change in membrane shape induced by protein adsorption and the corresponding change in membrane surface properties is not yet fully understood. This calls for a model that can describe such coupling. Furthermore, since asymmetry is one of the hallmarks of biological structures, it is important that such a model not be restricted to axisymmetric geometries. Here, we develop a two dimensional model  that can capture local inhomogeneities in the membrane properties in response to protein-adsorption. Using this 2D model, we show how the local tension of the membrane changes in response to protein adsorption and how this drives surface flow of the membrane's consituent lipids.

Biological membranes have unique mechanical properties. In an aqueous medium, lipids aggregate into quasi-2-dimensional bilayer sheets and adopt a configuration that minimizes the exposure of their hydrophobic parts. Even a plain lipid bilayer displays strange mechanical behaviors. In the plane of the membrane, it resembles a nearly incompressible viscous fluid, while in bending it behaves somewhat like an elastic solid. In-plane lipid flow has been observed in experimental systems such as tether formation and micropipette aspiration of membranes \cite{Fisher1992}. In cellular processes, the plasma membrane is thought to have a membrane `reservoir' that can act as a source of lipids \cite{Raucher1999}. This reservoir allows for lipid flow during dynamic events such as spreading, motility and endocytosis \cite{Raucher1999}. While the idea of lipid flow on the surface of membranes has been explored previously \cite{Secomb1982a,Secomb1982b}, most of these models focused only on lipid flow on a  cell or a vesicle of fixed shape. 

Simultaneous lipid flow and membrane shape change is a complex phenomenon. Hydrodynamics in membrane systems, in general, must include the viscosity of the surrounding bulk fluid \cite{Seifert1993,Fournier2009}, the in-plane flow of lipids  \cite{Arroyo2009,Rangamani2012,Rahimi2012, Seifert1993} and inter-monolayer friction  \cite{Fournier2009,Arroyo2009,Seifert1993,Rahimi2012}. As shown in \cite{Rahimi2012,Sens2004,Seifert1993}, each of these viscosities sets a different time scale for the membrane dynamics. In particular, for relaxation from fluctuations, the time scale set by bulk viscosity and inter-monolayer friction is important \cite{Seifert1993}. Here, we study the lipid flow on the membrane surface required to accommodate  shape changes  and surface area due to curvature induced by protein adsorption. For this, the relevant time scale for shape change and  curvature is set by the surface viscosity of the membrane \cite{Rahimi2012}.  

The paper is organized as follows:  In section \ref{sec: Viscous Elastic Model}, we propose  a 2D viscous-elastic formulation of lipid membrane dynamics that accounts for local spontaneous curvature due to protein adsorption. In section \ref{sec:Results}, using numerical simulations, we show that changes in the local spontaneous curvature due to protein adsorption alters the local tension. In turn, this drives lipid flow and consequent membrane shape change. In section \ref{sec:Discussion}, we elaborate on the  results from the simulations and their relevance to protein-adsorption phenomena.

\section{Viscous Elastic Model}
\label{sec: Viscous Elastic Model}
 
We use a Monge parametrization to describe the manifold representing the membrane. The position of a point on the membrane is given by
 
\begin{equation}
\label{eqn:position}
\rb = x_\alpha \eb_{\alpha} + z(x_1,x_2,t)\eb_3, \ \ \alpha \in \{1,2\},
\end{equation}

\begin{figure*}[h]
\includegraphics[scale=1]{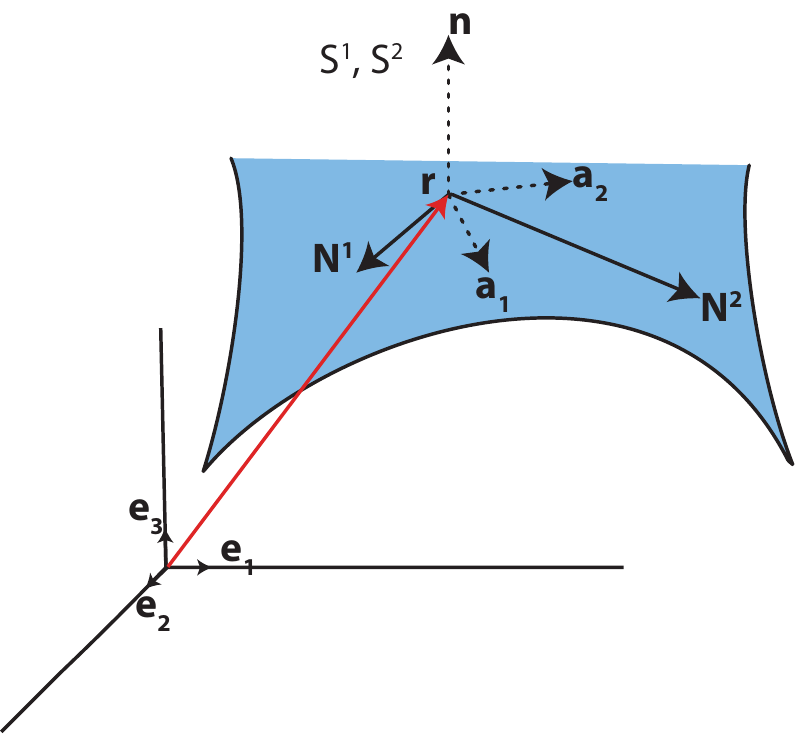}
\caption{The tangent basis $\ab_1$ and $\ab_2$ can be constructed at any point on the membrane as $\ab_\alpha=\rb_{,\alpha}$, where $\rb$ is the position given by Eq. (\ref{eqn:position}) and $\alpha\in\{1,2\}$. The surface normal $\nb$ is then $\frac{\ab_1\times\ab_2}{|\ab_1\times\ab_2|}$. The tangential components of the stress vectors $\Nb^1$ and $\Nb^2$ lie on the surface whereas the normal components $S^1$ and $S^2$ are along the surface normal. The total stress vectors are given by $\Tb^1=\Nb^1+S^1\nb$ and  $\Tb^2=\Nb^2+S^2\nb$. The definitions of  $\Nb^{\alpha}$ and $S^\alpha$ are given in \cite{Rangamani2012}.}
\label{fig:coordinates}
\end{figure*}

\noindent Here $\eb_1$, $\eb_2$ and $\eb_3$ are the orthonormal coordinate axes (see Figure~\ref{fig:coordinates}). The natural tangent bases on the membrane are $\ab_{\alpha} = \rb_{,\alpha} = \eb_{\alpha} + z_{,\alpha}\eb_3$, where $()_{,\alpha}$ represents the partial derivative with respect to $x^\alpha$. The components of the induced metric and curvature tensors in this parametrization are given by

\begin{eqnarray}
a_{\alpha\beta} &=& \ab_{\alpha}\cdot\ab_{\beta} = \delta_{\alpha\beta} + z_{,\alpha}z_{,\beta},\\
\ b_{\alpha\beta} &=& \nb \cdot \rb_{,\alpha\beta}= z_{,\alpha\beta}/\sqrt{a}.
\end{eqnarray}
respectively \cite{Sokolnikoff1964}. Here $a = det(a_{\alpha\beta})$, and $\nb = \frac{\eb_3 - z_{,\alpha}\eb_{\alpha}}{\sqrt{a}}$ is the membrane normal. Also, the dual metric and dual curvature components are given by $a^{\alpha\beta} = (a_{\alpha\beta})^{-1}$ and $b^{\alpha\beta} = a^{\alpha\lambda}a^{\beta\mu}b_{\lambda\mu}$, respectively.

Protein adsorption onto the membrane induces a change in the local curvature of the membrane. This can be  modeled by using a modified local form of the Helfrich energy density per unit area given by \cite{Helfrich1973}: 
 
\begin{equation}
\label{eqn:energy}
W =  k(H-C)^2+\bar{k}K,
\end{equation}

\noindent where $H=\frac{1}{2}a^{\alpha\beta}b_{\alpha\beta}$ and $K=\frac{1}{2}\varepsilon^{\alpha\beta}\varepsilon^{\lambda\mu}b_{\alpha\lambda}b_{\beta}{\mu}$ are the mean and Gaussian curvatures, respectively, and $k$ and $\bar{k}$ are the corresponding bending moduli. $C(x^1,x^2)$ is the spontaneous curvature induced by the proteins. (The difference between our form of the energy and the Helfrich energy is a factor of $\frac{1}{2}$, which is only a constant carried through all the calculations.)

With this machinery, we now develop a model for lipid bilayers with surface flow in the presence of protein-induced spontaneous curvature. 
The equations of motion in the absence
of inertia are simply the equations of mechanical
equilibrium. The force balance on the membrane, subjected to a net
lateral pressure $p$ in the direction of the local surface
unit normal $\nb$, can be summarized in the compact
form \cite{Steigmann1999}:
 
\begin{eqnarray}
\Tb^\alpha_{;\alpha}+p\nb={\bf 0}\\
\Tb^\alpha=\Nb^\alpha+S^\alpha\nb
\end{eqnarray}
where $\Tb^\alpha$ are the stress vectors, $\Nb^\alpha$ are tangential stress vector fields that are constitutively determined, and depend on the energy per unit mass of the membrane. $S^\alpha$ is a contravariant vector field that contains the normal component of the stress vector \cite{Rangamani2012,Steigmann1999}. 
We use a balance law formulation (see \cite{Rangamani2012}) to derive the equations of motion for a membrane with intra-surface flow in the presence of curvature inducing proteins. This approach has two distinct advantages over the commonly used global energy minimization approach. First, the local stress balance approach allows us to account for local inhomogeneities in the membrane (see \cite{Steigmann1999} for a detailed explanation). Second, the local force balance allows us to include dissipation arising from intra-membrane viscosity; the viscous stress contributes to $\Nb^\alpha$ only, i.e. including in-plane flow of lipids affects only the tangential stress terms \cite{Rangamani2012,Rahimi2012}.

The velocity of any point $\rb$ on the membrane may be decomposed into  tangent and normal components as
 
\begin{equation}
\dot{\rb} = v^\alpha\ab_{\alpha} + w \nb, 
\end{equation}
where $v^{\alpha}$ are the intra-membrane flow velocities and $w = z_{,t}/\sqrt{a}$ is the normal velocity of the membrane. 
The membrane is assumed to be incompressible, which results in the following constraint \cite{Rangamani2012}:
\begin{equation}\label{eqn:compressibility} 
v_{;\alpha }^{\alpha}-2wH=0.
\end{equation}
The incompressibility constraint is implemented using a Lagrange multiplier $\gamma$ \cite{Rangamani2012,Jenkins1977}. 

Following the procedure in \cite{Rangamani2012}, the modified normal and intra-membrane 
flow equations of motion in the absence of inertia including the spontaneous curvature are
 
\begin{eqnarray}
\label{eqn:viscous_shape}
\nonumber
k[\Delta (H-C)+2(H-C)(H^2+HC-K)]&&\\-2\lambda H 
+2\nu \lbrack b^{\alpha \beta }d_{\alpha
\beta }-w(4H^{2}-2K)]&=&p,
\end{eqnarray}
and
 
\begin{equation}
\label{eqn:viscous_flow}
\lambda _{,\gamma }-4\nu wH_{,\gamma }+2\nu (a^{\alpha \mu }d_{\gamma \mu
;\alpha }-w_{,\alpha }b_{\gamma }^{\alpha })= 2k(H-C)\frac{\partial C}{\partial x^{\gamma}} ,
\end{equation} 
where $\Delta(\cdot)=(\cdot)_{;\alpha\beta}a^{\alpha\beta}$ is the surface Laplacian, $\nu$ is the membrane viscosity, $\lambda=-(W+\gamma)$ and 
 
\begin{eqnarray}
d_{\alpha\beta} &=& \frac{1}{2}(v_{\alpha;\beta} + v_{\beta;\mu}) \ , \\
v_{\alpha;\beta} &=& v_{\alpha,\beta} - v_{\phi}\Gamma^{\phi}_{\alpha\beta}.
\end{eqnarray}
Here, $v_{\alpha}$ are the covariant components of the velocity vector, $()_{;\alpha}$ is the covariant derivative and $\Gamma^{\phi}_{\alpha\beta}=z_{\,\phi}z_{,\alpha\beta}/a$ are the Christoffel symbols \cite{Sokolnikoff1964}. 

Changes in the membrane shape in response to protein adsorption are obtained by solving  Eqs.~(\ref{eqn:viscous_shape}), (\ref{eqn:viscous_flow}) and (\ref{eqn:compressibility}) along with appropriate boundary conditions \cite{Rangamani2012}. 
 
\subsection{Reduction to an Elastic Model and Interpretation of Surface Tension}
\label{sec: Elastic Model}
 
When the membrane viscosity, $\nu$, is zero, the above model reduces to the elastic model of bilayer membranes.  Eq. (\ref{eqn:viscous_shape}) reduces to the well-known shape equation for the Helfrich energy \cite{Steigmann1999}
 
\begin{eqnarray}
\label{eqn:elastic_shape}
k[\Delta (H-C)+2(H-C)(H^2+HC-K)]-2\lambda H=p.
\end{eqnarray}
The spatial variation of $\lambda$ is given by
 
\begin{equation}
\label{eqn:elastic_flow}
\lambda _{,\gamma }= -\frac{\partial W}{\partial x^{\gamma}}{\mid_{\exp}} =2k(H-C)\frac{\partial C}{\partial x^{\gamma}}.
\end{equation}

It should be noted that the Lagrange multiplier $\gamma$, used to implement the area incompressibility constraint (Eq.~(\ref{eqn:compressibility})), is defined as the surface pressure of the membrane \cite{Rangamani2012}. This interpretation is consistent with the notation used in \cite{Jenkins1977}. In the absence of bending and spontaneous curvature, the force normal to any curve is given by $f_\nu=\lambda=-\gamma$ (see \cite{Rangamani2012}), and therefore $\lambda$ can be understood as the tension in a flat membrane. Furthermore, in the special case of zero spontaneous curvature and non-zero mean curvature, $\lambda=constant$ everywhere, see Eq.~(\ref{eqn:elastic_flow}). This constant value of $\lambda$ must be provided as an input parameter to solve the system of equations \cite{Steigmann1999}, and is widely interpreted to be surface tension in the literature \cite{Derenyi2002}. 

Surface tension can be estimated from studying vesicles \emph{in vitro} using different approaches: (i) capillary methods, (ii) fluctuation spectra of vesicles and (iii) membrane tethering experiments. In capillary methods, the vesicle diameter and the osmotic pressure difference between the exterior and the interior of the vesicle can be controlled. Since the vesicle is a sphere and of constant mean and Gaussian curvatures, we can use the Young-Laplace equation resulting from the simplifications of Eq.~(\ref{eqn:elastic_shape}) to obtain $\lambda$ (See chapter 11 of \cite{Phillips2009} for detailed explanation). Estimates of surface tension can also be obtained from analysis of the fluctuation spectra of giant unilamellar vesicles using high resolution contour detection techniques \cite{Pecreaux2004}. Alternatively, in the case of membrane tethering experiments, one may pull tubes and measure the force vs. length of the tubule curves and obtain surface tension as parameter fits to the theory in \cite{Derenyi2002}\footnote{Instead of solving problems using $\lambda$, one can also work in the space of surface pressure $\gamma$. If $\gamma$ is used as a dependent variable, the Gaussian modulus enters into Eqs.~(\ref{eqn:viscous_shape},\ref{eqn:viscous_flow}). However, it is well known that Gaussian modulus measurements vary quite widely in experiments \cite{Siegel2004,Siegel2006} and can only be calculated accurately from simulations \cite{Hu2012}. Using $\lambda$ is beneficial because it can be inferred from experiments on large spherical vesicles, where the force normal to any boundary is $f_\nu = \lambda = \frac{p}{2H}$, where $p$ is the pressure difference and $H$ is the mean curvature. Hence, this value of $\lambda$ can be used as a boundary condition to solve Eq.~(\ref{eqn:viscous_flow}).}.

  However, in general, $\lambda$ is not a uniform field and has to be solved for simultaneously with the shape of the membrane using the above model. Since $\lambda$ is non-homogeneous and depends on the local inhomogeneities, herein we refer to it as the \emph{local tension} in the membrane.   

\subsection{Limitations of the Elastic Model}
There are two equations for a single unknown $\lambda$ (Eq.~(\ref{eqn:elastic_flow})), posing a numerical challenge for solving $\lambda$ and the shape simultaneously. 
\noindent 
As noted earlier the equations of motion tangential to the membrane, Eq. \eqref{eqn:elastic_flow}, yield the spatial dependence of the local tension, $\lambda$, in the elastic model. Solving Eq. \eqref{eqn:elastic_flow}, the value of the local tension can be obtained as 
\begin{equation}
\label{eqn:elastic_flow_integralform}
\lambda = \int_{x^1_0}^{x^1} 2k(H-C)\frac{\partial C}{\partial x^{1}} dx^1+ \int_{x^2_0}^{x^2} 2k(H-C)\frac{\partial C}{\partial x^{2}}dx^2 + \lambda(x^1_0,x^2_0),
\end{equation}
\noindent where $(x^1_0,x^2_0)$ correspond to a point where the local tension is known, for example, any point on the boundary. Using Eq.~\eqref{eqn:elastic_flow_integralform}, the shape equation Eq. (\ref{eqn:elastic_shape}) can be reduced to  
\begin{equation}
\begin{split}
\label{eqn:elastic_shape_integrodifferential}
k[\Delta & (H-C)+2(H-C)(H^2+HC-K)]-  \\
 &2 \Bigg(\int_{x^1_0}^{x^1} 2k(H-C)\frac{\partial C}{\partial x^{1}}dx^1 \\
 & + \int_{x^2_0}^{x^2} 2k(H-C)\frac{\partial C}{\partial x^{2}}dx^2 + \lambda(x^1_0,x^2_0) \Bigg) H=p.
\end{split}
\end{equation}
Eq.~\eqref{eqn:elastic_shape_integrodifferential} is a complicated integro-differential equation that is difficult to solve both analytically and numerically. Another way to solve the equations in the elastic model (Eq.~(\ref{eqn:elastic_shape}) and Eq.~(\ref{eqn:elastic_flow})) numerically is to use vanishing viscosity methods, wherein dynamics is added to the system by means of an artificial or numerical viscosity and the solutions are obtained by taking the limit as viscosity tends to zero \cite{Dafermos,Bianchini2005,Chen2010}. The equations that result using this method are similar to the viscous-elastic model Eq.~\eqref{eqn:viscous_flow} with differences primarily in the viscosity matrices that need to be constructed. Moreover, one needs to solve the system of equations repeatedly under the limit viscosity tends to zero. However, it is known that bilayers have in-plane viscous behavior, and if we use the viscous-elastic model presented in Section~\ref{sec: Viscous Elastic Model}, viscosity naturally enters into the system of equations. These equations can then be solved using methods available for solving the regular Navier-Stokes equations. Therefore, the viscous-elastic model provides a physically meaningful way to compute the effects of proteins on membrane shape, and thereby the changes in the local surface properties of the membrane.


However, if the system is axisymmetric, then the solution can be easily obtained because one (curvilinear) coordinate specifies the system at equilibrium (as shown in \cite{Lipowsky2012,Rahimi2012,Agrawal2008}). The viscous-elastic model (Eqs.~(\ref{eqn:viscous_shape}), (\ref{eqn:viscous_flow}) and (\ref{eqn:compressibility}) on the other hand, allows us to simultaneously solve for membrane shape, lipid flow and $\lambda$ in both axisymmetric and 2D configurations. In the following section, we will demonstrate this with some numerical examples. 

\subsection{Spontaneous Tension as a Special Case} \label{sec:Lip_spont_tens}

Here, we compare our work to the notion of `spontaneous tension' introduced in \cite{Lipowsky2012} to model proteins or solute particles that adsorb onto the lipid membrane and show that it is a special case of the elastic model presented in Section~\ref{sec: Elastic Model}. In \cite{Lipowsky2012}, the effect of adsorbed proteins is modeled by means of a spontaneous curvature, as in this work, where the total energy is given by \cite{Helfrich1973}
\begin{equation}\label{eqn:energy}
\mathcal{W} = \int W \, \mathrm{d} A = \int( k(H-C)^2+\bar{k}K \, )\mathrm{d} A  .
\end{equation} 
As assumed in \cite{Lipowsky2012}, consider the case when the resultant mean and Gaussian curvatures are significantly smaller than the applied spontaneous curvature in a certain area $A_c$: 
\begin{equation}\label{eqn:LipowskyCond}
H \ll C, \
K \ll C^2  .
\end{equation}
Here, the total energy in the area $A_c$ is given by
\begin{equation}\label{eqn:energy}
\mathcal{W} \approx \int_{A_c} kC^2 \, \mathrm{d} A = kC^2 A_c . 
\end{equation} 
Therefore, the effect of spontaneous curvature can be understood as an apparent constant tension $\lambda_{sp}$ over the area $A_c$ given by 
\begin{equation}\label{eqn:spontens}
\lambda_{sp} = \frac{\delta \mathcal{W}}{\delta A_c} = kC^2.  
\end{equation}
Eq.~(\ref{eqn:spontens}) is the central result in \cite{Lipowsky2012}. 

In the elastic model limit presented in our work, specializing to the case where \eqref{eqn:LipowskyCond} is true, we have from \eqref{eqn:elastic_flow}
\begin{equation}
\label{eqn:elastic_flow-1}
\lambda _{,\gamma } \approx 2k(-C)\frac{\partial C}{\partial x^{\gamma}} = -(kC^2)_{,\gamma}.  
\end{equation} 
Integrating Eq. \eqref{eqn:elastic_flow-1} gives
\begin{equation}\label{eqn:elastic_flow-2}
\lambda(x^1,x^2) - \lambda(x^{1}_0,x^{2}_0) = - kC^2 , 
\end{equation}
where $(x^{1}_0,x^{2}_0)$ is any point outside the protein patch. This tells us that the local tension differs between the protein patch and the rest of the membrane. This value is the spontaneous tension $kC^2$ as in \eqref{eqn:spontens}. Hence, the result in \cite{Lipowsky2012} is a special case of the present theory under the assumptions \eqref{eqn:LipowskyCond}. Finally, from the general viscous-elastic model, we find that the conditions \eqref{eqn:LipowskyCond} do not hold in general as will be discussed in the following section. 

%

\section{Results}
\label{sec:Results}

\subsection{Effects of protein-induced spontaneous curvature}
 
In this work, we assume that the proteins adsorb onto a small patch on the membrane and do not diffuse \cite{Agrawal2008}. There are many protein complexes that bind to the membrane at specific sites and do not disperse \cite{Karotki2011,Stachowiak2012, Kishimoto2011, Walther2006}. 
The total system we study is a square of side $1000$ nm, with a preferred protein binding patch of side $50$ nm in the center. The spontaneous curvature $C(x^1,x^2,t)$ generated by the proteins is assumed to be approximately uniform in the square patch and modeled as
\begin{eqnarray}
\nonumber
C(x^1, x^2,t)&=&\frac{1}{4}C_0(t)\left[\tanh(x^1-25)-\tanh(x^1+25)\right]\\
&&\left[\tanh(x^2-25)-\tanh(x^2+25)\right],
\end{eqnarray}
where $C_0(t)$ is the magnitude of the spontaneous curvature, and depends on the density of proteins \cite{Stachowiak2012}.
The kinematic boundary conditions for the membrane position along the edges of the square patch are
\begin{equation}
z=0\quad \text{and}\qquad \nb=\eb_3 
\end{equation}
We model the square patch as two closed (top and bottom) boundaries with no-slip boundary conditions. The left and right boundaries are open and allow for lipid flow based on the following traction boundary conditions: 
\begin{equation}\label{eqn:BCs} 
 \frac{1}{4}kz_{,11}^2=\nu v_{1;1}\qquad \text{and} \qquad v_{2;1}+ v_{1;2}=0.
\end{equation}
based on the general expressions for the edge forces \cite{Rangamani2012}. Additionally, the value of $\lambda$ is specified on the open boundaries as $\lambda_0=5\times 10^{-4}$ pN/nm  \cite{Lipowsky2012,Pecreaux2004,Faris2009}. The other parameters used in the model are membrane bending rigidity $k= 82$ pN$\cdot$nm \cite{Lipowsky1995} and intra-membrane viscosity $\nu=1 \times 10^{-4}$ pN$\cdot$s/nm \cite{Hochmuth1987}.

Protein adsorption in the center of the patch, and therefore $C_0$, is modeled as a linear function of time as shown in Figure~\ref{fig:position}(A). This function captures the increasing density of proteins that are absorbed onto the membrane over time. The partial differential equations are solved using finite element methods \cite{Zien2000,Hughes1986,bathe,babuska} in COMSOL Multiphysics\textsuperscript{\textregistered}. Note that all variables appear as second derivatives in space, while $\lambda$ alone appears as a first derivative. In order to achieve numerical convergence, we use a linear shape function for $\lambda$ and quadratic shape functions for the other variables \cite{bathe,Hughes1986,babuska}.

In response to protein adsorption, the shape of the membrane changes with time as shown in Figure~\ref{fig:position}(B). The height of the deformed membrane in the center is shown in Figure~\ref{fig:position}(A). Increasing the spontaneous curvature increases the surface area. To accommodate this change, lipids flow in from the open boundaries. Once the spontaneous curvature attains a steady value, the height of the membrane also attains a steady value as expected. The shape of the membrane in response to protein adsorption (Figure~\ref{fig:position}(B)) resembles the early endocytic invagination due to coat protein adsorption (Figure~\ref{fig:position}(C)) \cite{Buser2013}.

$\lambda$ is a non-linear function of the spontaneous curvature $C$ (see Figures~\ref{fig:lambda}(A) and \ref{fig:position}(A)). Analysis of the local tension profile along the membrane shows that for small values of spontaneous curvature, there is no measurable inhomogeneity in the value of the local tension $\lambda$. As the spontaneous curvature increases, the value of lambda decreases only in the protein patch (Figure~\ref{fig:lambda}(B)). 

\subsection{Interaction between two protein patches}
As seen in Figures \ref{fig:position} and \ref{fig:lambda}, protein adsorption leads to a global change in membrane shape and a local change in $\lambda$. If one patch of adsorbed proteins causes a local change in $\lambda$, then how do two separate regions of proteins interact with one another? In other words, does the change in $\lambda$ remain confined to regions with adsorbed proteins or does it propagate across regions with no proteins? To understand this interaction, we performed simulations with two patches of membranes that are placed at different distances apart from one another and studied the evolution of shape and local tension in these systems.

We carried out simulations in a square domain of side $3000$ nm for two protein patches, each a square of side $50$ nm, separated by different distances. As shown in Figure~\ref{fig:2_patches_Z}(A(i)), for two protein patches separated by a distance smaller than the characteristic decay length ($0$, $50$, $100$ nm), the curvature in the region between them is affected. The region in the center does not go to $z=0$ but takes on an intermediate height between the two maxima  (Figure~\ref{fig:2_patches_Z}(B)). When the separation distance is $600$ nm, the two patches continue to influence each other. This can be inferred by the fact that the mean curvature does not go to zero in the region between the patches (Figure~\ref{fig:2_patches_Z}(A(ii))). 
However, when the separation distance is $2000$ nm, the mean curvature between the two patches goes to zero (Figure~\ref{fig:2_patches_Z}(A(ii))). A magnified view of the mean curvature between these two patches is also shown in (Figure~\ref{fig:2_patches_Z}(A(ii)), bottom panel). This separation distance is much greater than the characteristic decay length $L_d=\sqrt{\frac{k}{2\lambda}} \approx 280 \ nm$ obtained from small perturbations to a flat bilayer in \cite{Derenyi2002} (See Eq.~(9) in \cite{Derenyi2002}).

In contrast, $\lambda$ remains a localized effect; in the region of protein adsorption $\lambda$ attains the same value as it would in a single patch (Figure~\ref{fig:2_patches_lambda}). In the region between the two protein patches, $\lambda$ remains $\lambda_0$. This effect can be explained in part by studying Eq.~(\ref{eqn:elastic_flow}), where the change in $\lambda$ is given by the first-order partial differential equation. Therefore, in the region where there are no proteins, $\partial C/\partial x^\alpha$ goes to zero, and $\lambda$ is given the value at the boundary $\lambda_0$. When the distance between the patches is decreased  (Figure~\ref{fig:2_patches_lambda}), the effect on curvature is even more pronounced while $\lambda$ continues to be a local effect. When the two regions share a boundary (Figures~\ref{fig:2_patches_Z}(A,B), \ref{fig:2_patches_lambda}), the effect of shape and $\lambda$ is essentially the same as simulating a single rectangular patch. 

From these results, we can infer that for these boundary conditions, the value of $\lambda$ depends only on $C_0$ and not on the size, location and shape of the patch. Thus, protein adsorption creates not only changes in membrane curvature at long length scales but also changes to membrane tension locally. Moreover, two different protein patches can be considered completely independent when they are separated by distances larger than the characteristic decay length $L_d=\sqrt{\frac{k}{2\lambda}}$. Finally, as mentioned before in Section~\ref{sec:Lip_spont_tens}, the value of H in the membranes are not small compared to the imposed spontaneous curvature as seen in Figure~\ref{fig:2_patches_Z}(A)  thus violating the conditions Eq.~\ref{eqn:LipowskyCond} as assumed in \cite{Lipowsky2012}. Since our model does not make such assumptions, it allows us to calculate the changes in $\lambda$ in the most general cases.



\section{Conclusions and Discussion}
\label{sec:Discussion}
Biological membranes have unique mechanical properties. In the plane of the membrane, they resemble a nearly incompressible viscous fluid, while in bending they behave somewhat like an elastic shell. The extent to which this dichotomy holds depend on the length and time scales one investigates. Molecular and Brownian dynamics simulations can capture membrane dynamics at the smallest scale, but with current tools it is still daunting to follow and to extract their collective properties. Thus a larger scale model  is desirable that still captures the mechanical properties. The search for such a model, however, has proven more difficult than would first appear. 

An attractive theoretical construct is to model a bilayer as a 2-dimensional differentiable manifold and then to endow this surface with the appropriate mechanical and material properties. Helfrich proposed such a model which treats the manifold as an elastic shell whose bending behavior is captured by an energy density functional that depends only on the manifold's local curvatures \cite{Helfrich1973}. Models of this sort proved sufficient for radii of curvatures much larger than the membrane thickness, and rich geometric behavior could be obtained by minimizing the elastic bending energy. 

The situation becomes more subtle when one attempts to use the manifold model to examine membranes that exhibit larger curvatures due to mixtures of lipid types or to interactions with proteins that bind to one leaflet or the other. This brings into play properties tangential to the membrane surfaces. A lipid bilayer has two separate interfaces, whereas the simple manifold model has but one. Still, one can introduce the notion of an `intrinsic' curvature due to, say, different lipids or proteins on each leaflet, and thus rescue the Helfrich model by including this `built in' curvature. It is, however, important to remember that the membrane is held together not by attractive forces between the lipids, but by the interfacial tension developed by the aqueous solvent in which the lipids are dissolved. That is, the hydrogen bond network between the water molecules provides the only `tensile' force in the system. Thus anything that disturbs the hydrogen bond network on one interface will introduce a bending moment that arises from `outside' the membrane proper. Introducing a `spontaneous curvature' only mimics this external force in certain aspects. In particular, when the membrane curves due to the spontaneous curvature term in the energy expression, it does so not by generating a bending moment: the bending is `kinematic', and does not discriminate between, say, an absorbed protein on one leaflet and the action of a lipase on the other face.

This limitation of the manifold elastic model becomes more acute if one wants to model the fluid properties of the membrane. Individual lipids diffuse rapidly in the plane of the membrane, and in certain situations flip back and forth between the monolayers. The elastic manifold  model cannot handle surface diffusion and convection directly, and so one must introduce additional structure. To model surface flows one can define a velocity field on the manifold that responds to gradients in surface tension or surface pressure. But surface tension is a body force imposed on the membrane by the solvent. To drive a flow we must introduce a driving force, along with gradients in surface lipid density in order to handle convection. The manifold must bear all these artifices to avoid treating the bilayer for what it is: a collection of particles held together by a solvent.

\begin{figure*}[h]
\includegraphics[scale=0.75]{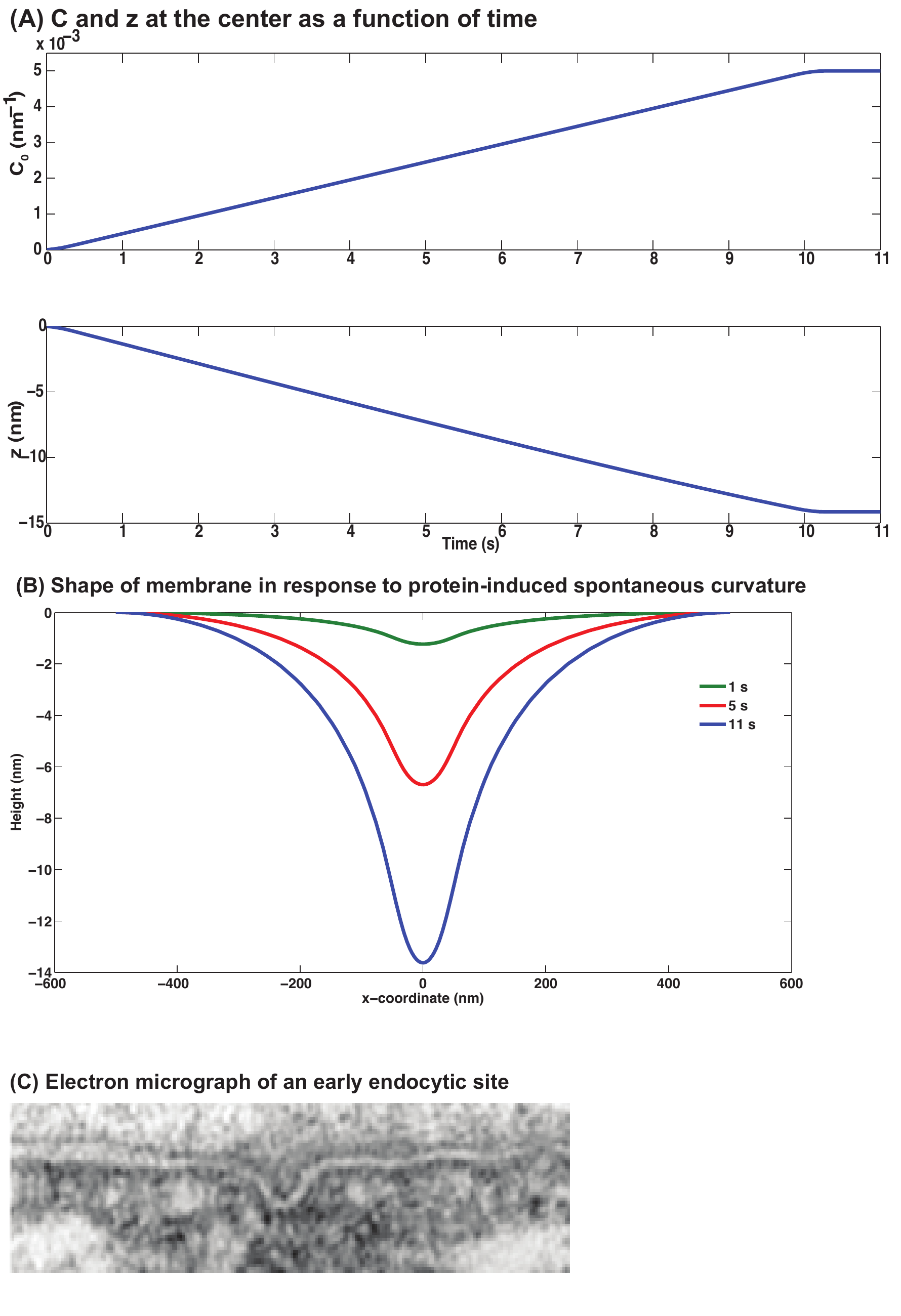}
\caption{The time-dependent change in membrane shape due to protein adsorption. (A) Protein-induced spontaneous curvature C and height of the membrane $z$ in the center of the domain as a function of time. (B) Shape of the membrane corresponding to line $x^2=0$ at three different times. (C) An electron micrograph showing an early endocytic invagination \cite{Buser2013}.}
\label{fig:position}
\end{figure*}

\begin{figure*}[h]
\includegraphics[scale=1]{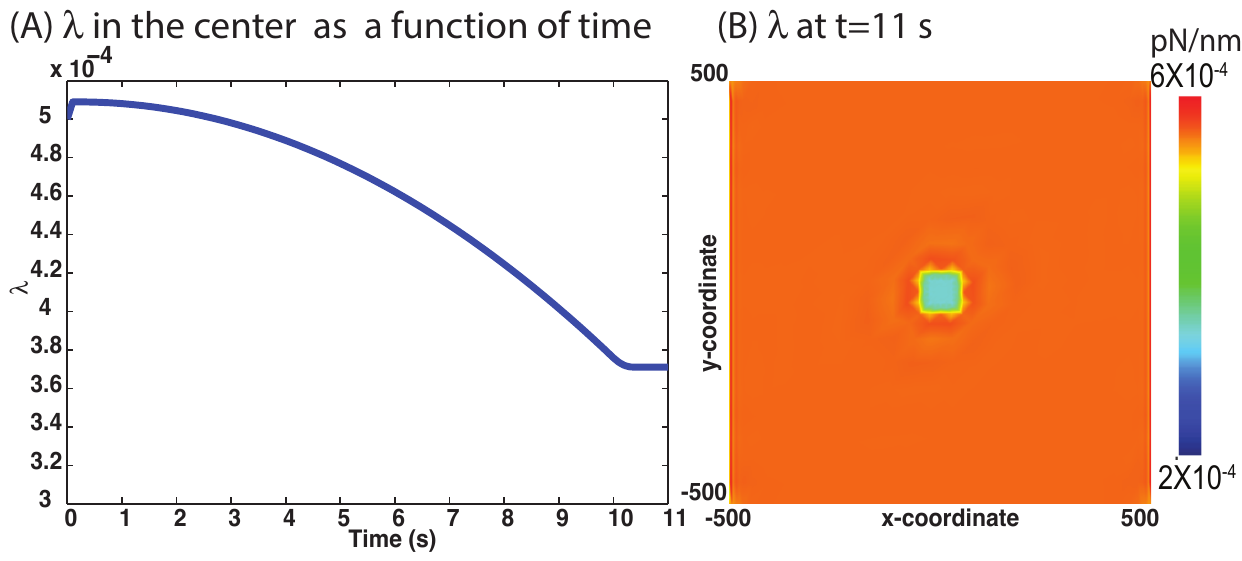}
\caption{Local tension variation on the membrane in response to protein adsorption. (A) $\lambda$ versus time in the center of the domain in response to protein adsorption. (B) Surface profile of $\lambda$ at 11 s.}
\label{fig:lambda}
\end{figure*}

We have shown here that a viscous-elastic description of the membrane is required to capture the local changes in membrane tension due to protein adsorption. In two dimensions, the Helfrich model results in a integro-differential equation that is hard to solve when the membrane is inhomogeneous. Introduction of membrane surface viscosity allows us to solve for the change in shape and local tension simultaneously in the case of an inhomogeneous membrane. This situation is of particular biological relevance: there are many cases where protein binding or membrane crowding is localized \cite{Kishimoto2011,Buser2013,Walther2006,Karotki2011}. The adsorption of proteins to specific membrane microdomains is often the first step in a biological process like endocytosis. In many of these process, the local regions where proteins bind are also regions where further shape changes take place. For example, in endocytosis, the early invagination is followed by tubulation and subsequent scission of the vesicle. In Figure \ref{fig:position}, we show how the spontaneous curvature of the proteins adsorbing on the membrane leads to an change in the shape of the membrane that is similar is size and shape to that of an early endocytic invagination. Our model and simulations also show that the change in shape is accompanied by a change in the local tension of the membrane in the local region where proteins are bound (Figure \ref{fig:lambda}). It is possible that the lowering of local tension of the membrane will have a functional impact on downstream effects of protein binding. In fact, one experimentally testable prediction from our work is that protein adsorption on the membrane can lower the local tension and the energy barrier for subsequent events in that region. If this is indeed the case, then the protein coated membrane will deform more for the same applied force when compared to the uncoated bare membrane. 

\begin{figure*}[h]
\includegraphics[width=\textwidth]{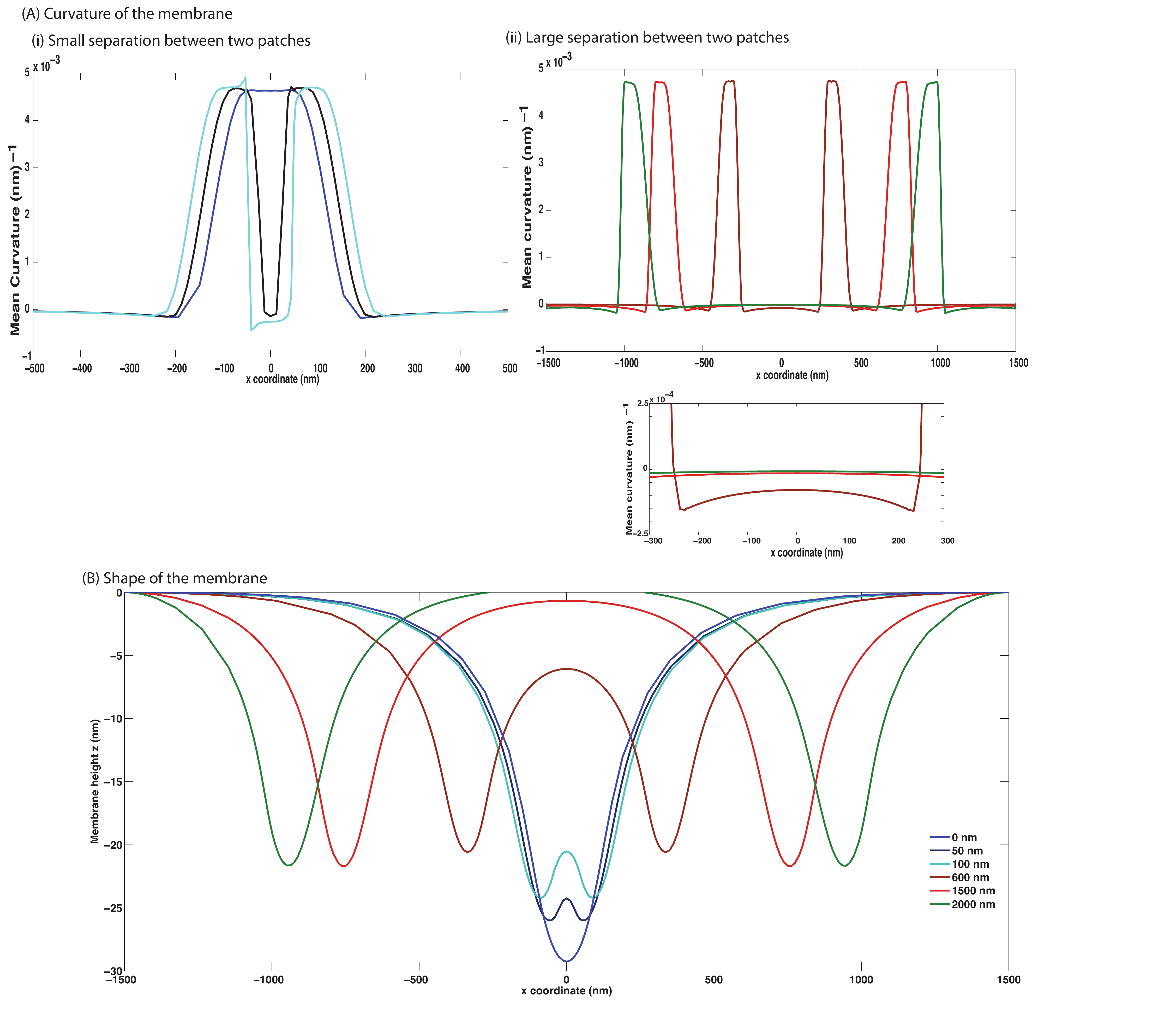}
\caption{Protein adsorption on two patches. (A) Membrane curvature $H$ corresponding to line $x^2=0$ (i) for a small separation ($L<L_d$) between two patches and (ii) for a large separation ($L>L_d$) between two patches. The bottom plot shows a magnified view of mean curvature $H$ between the two patches. (B) Membrane height $z$ for two patches with an edge-to-edge distance of 2000 nm, 1500 nm, 600 nm, 100 nm, 50 nm and 0 nm apart. As expected, when the two patches share a boundary, $H$ and $z$ respond as if a single large patch has proteins adsorbing on it. }
\label{fig:2_patches_Z} 
\end{figure*}

\begin{figure*}[h]
\includegraphics[width=\textwidth]{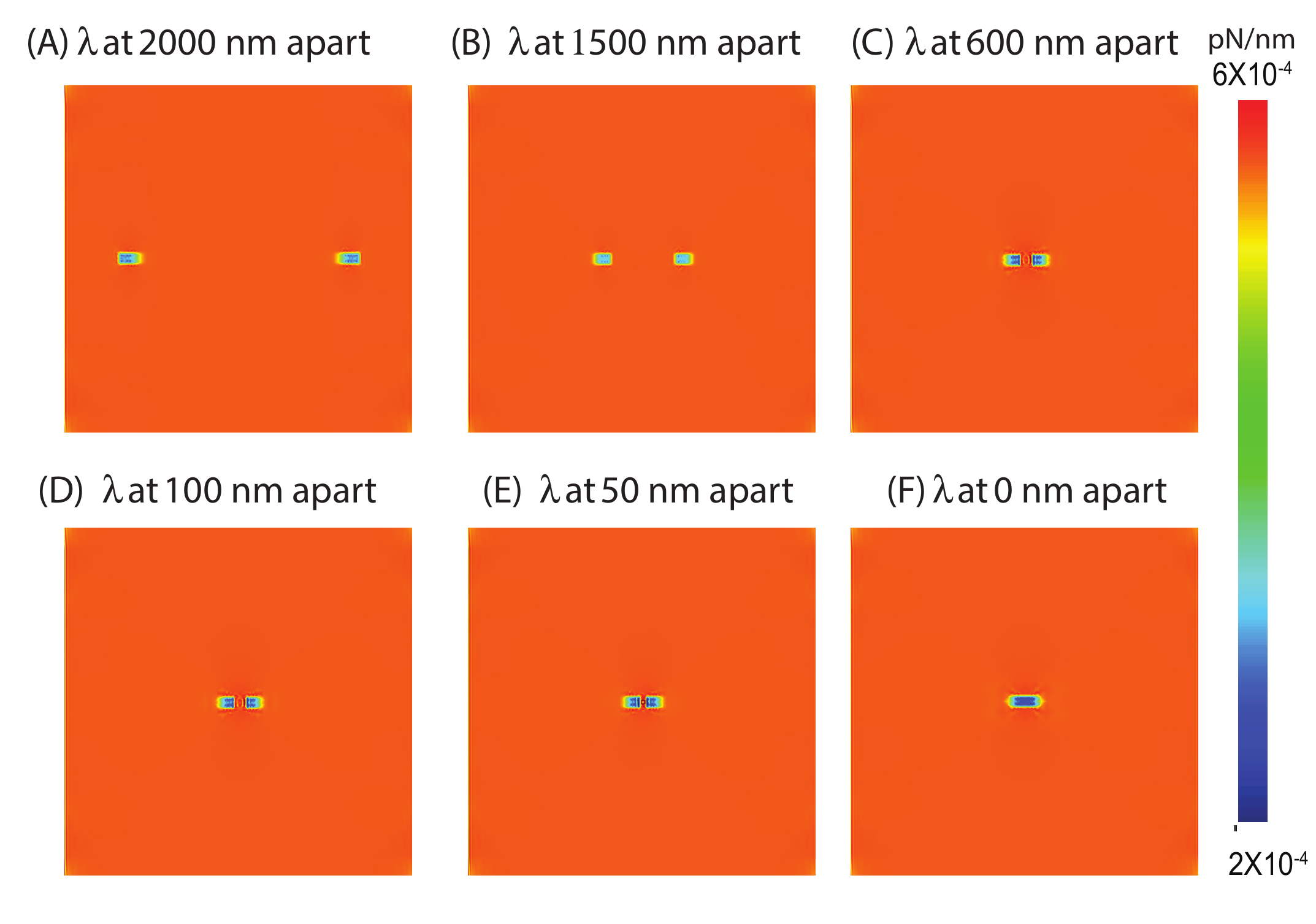}
\caption{ Local tension $\lambda$ for two patches with an edge-to-edge distance of (A) 2000 nm, (B) 1500 nm, (C) 600 nm, (D) 100 nm, (E) 50 nm and (F) 0 nm apart. The change in $\lambda$ remains confined to the local patch in all cases and when the two patches share a boundary, $\lambda$ behaves as if a single large patch has proteins adsorbing on it. }
\label{fig:2_patches_lambda} 
\end{figure*}

Another important result from our work is that when proteins can bind to two patches on the membrane, there is characteristic decay length that is given by the ratio of out-of-plane bending to the in-plane tension. When the two patches are located at less than or of the same order of magnitude as the decay length, the region between the two patches experiences the curvature changes even though local tension changes are always limited to the patch (Figures \ref{fig:2_patches_Z}, \ref{fig:2_patches_lambda}). When the two patches are much further away, then the curvature between the two patches relaxes to zero in the region between them. Thus, even though the local tension changes remain confined to the patch, the curvature effects are dependent on the characteristic decay length. 

It is also likely that, in addition to lipid flow, inclusion of lipid and protein diffusion will further impact the surface properties of the membrane. It should also be noted that proteins can induce both spontaneous mean and Gaussian curvature. Even if we were to include a spontaneous Gaussian curvature (similar to \cite{Seguin2012,Kim1998}) in the energy (Eq.~(\ref{eqn:energy})), a viscous-elastic formulation would still be necessary to describe the system uniquely. Including inter-monolayer friction and bulk liquid viscosity \cite{Fournier2009, Seifert1993}, along with membrane surface viscosity may also be necessary to accurately model membrane behavior in response to local inhomogeneities. These issues will be addressed in a future study. 


\newpage

\end{document}